# The Self-Organizing Society: A Grower's Guide

John E. Stewart[1]

***Abstract***: Can a human society be constrained in such a way that self-organization will thereafter tend to produce outcomes that advance the goals of the society? Such a society would be self-organizing in the sense that individuals who pursue only their own interests would none-the-less act in the interests of the society as a whole, irrespective of any intention to do so. This paper identifies the conditions that must be met if such a self-organizing society is to emerge. It demonstrates that the key enabling requirement for a self-organizing society is 'consequence-capture'. Broadly this means that all agents in the society must capture sufficient of the benefits (and harms) that are produced by their actions on the goals of the society. 'Consequence-capture' can be organized in a society by appropriate management (systems of evolvable constraints) that suppress free riders and support pro-social actions. In human societies these constraints include institutions such as systems of governance and social norms. The paper identifies ways of organizing societies so that effective governance will also self-organize. This will produce a fully self-organizing society in which the interests of all agents (including individuals, associations, firms, multi-national corporations, political organizations, institutions and governments) are aligned with the interests of the society as a whole.

[1] Evolution, Complexity and Cognition group, Center Leo Apostel, Vrije Universiteit Brussel, Krijgskundestraat 33, B-1160 Brussels, Belgium.
Email address: future.evolution@gmail.com (John Stewart)

# 1. Introduction

Is it possible to constrain a society in such a way that self-organization within the society will always tend to produce outcomes that further the collective goals of the society? Consider a society which has a primary goal of providing some conception of ‘the good life’ for its citizens, and maintaining this in the face of internal and external challenges. Is there a way to constrain the society so that actions that produce ‘the good life’ will tend to self-organize within the society, and contrary actions will not? It is worth noting here that in order for such a society to be fully self-organizing in this sense, the constraints that organize it would also have to be self-organizing.

Before addressing this question, it is necessary to clarify the use of the term ‘self-organize’ in this context. ‘Self-organization’ as a concept has been used in different ways in different contexts and fields, often with little understanding of what it entails. But my usage here is broadly consistent with the approach taken in the study of Complex Adaptive Systems. In that field, self-organization is generally said to occur when larger-scale order or organization is produced by the interactions of agents that pursue only their own immediate interests [1, 2]. This self-organization occurs even though the agents may have no intention of producing it. Furthermore, the wider-scale organization that emerges is not produced by the intervention of some centralized controller that plans the specific outcomes it wants and then specifies the particular behaviours that are necessary to produce these outcomes.

Consistent with this usage of the term ‘self-organization’, market-based economic systems in human societies are often given as examples of self-organizing processes. They are typically contrasted with centrally-planned command economies [3]. In an appropriately-governed economic market, the pursuit by individuals of their immediate interests can benefit other participants in the market, irrespective of whether the individuals have any intention or desire to do so. This form of self-organization in human economic systems is often referred to as an ‘invisible hand’ process [4].

Drawing on these understandings of ‘self-organization’, the term ‘self-organizing society’ is used here to refer to a society in which the pursuit by members of the society of their immediate self-interest would none-the-less also advance the collective interests and goals of the society, whatever they may be. It is worth emphasizing here that in order to qualify as a self-organizing society, this condition would also have to apply to members of the society who are involved in establishing and adapting governance and other institutions. In other words, the pursuit by these members of the society of their immediate interests would establish institutions that advance the interests and goals of the society as a whole.

Using the terminology of the ‘invisible hand’, a self-organizing society is one in which invisible hand processes are not restricted to the functioning of appropriately-managed markets. In a self-organizing society, invisible hand processes would operate in all domains, ensuring for example that institutions would also serve the common good (including by managing markets appropriately). In other words, the satisfaction of needs and wants that cannot be met by economic markets would none-the-less be satisfied by other invisible hand processes.

It is clear that modern human societies are not self-organizing societies in the sense used here: outcomes that most would consider to be socially undesirable are often produced in modern societies when their own immediate interests are pursued by individuals, businesses, corporations, political organizations and other agents. For example:

(a) Far from serving only the common good, economic markets also self-organize undesirable outcomes including financial crises and environmental damage, even though no market participants may intend to produce these outcomes.
(b) Powerful interests influence governments to act in ways that advance private interests at the expense of outcomes that would be more beneficial for the society.
(c) Significant proportions of humans live and die without their fundamental needs and desires being met effectively or fairly. Nor do they have the opportunity to develop fully their potential to contribute positively to society.
(d) The pursuit by various groups of their individual interests can produce wars and conditions that facilitate wars, including the possibility of nuclear annihilation.
(e) Large-scale criminal activities emerge and exploit opportunities to advance the interests of their participants at the expense of societal interests through theft, manipulation, fraud, etc.
(f) Corporations and other business interests are able to find many ways to advance their interests without producing goods and services that efficiently satisfy the genuine needs and desires of others, including through dishonesty, monopoly power, tax avoidance, anti-competitive practices, the manipulation and distortion of desires through advertising, etc.
(g) Politicians, political parties, bureaucrats, governmental bodies and others involved in establishing governance in modern societies find many ways to advance their own interests in ways that do not benefit the society.

The fact that typical modern societies are not self-organizing can also be demonstrated by showing that they lack key features that would arise in self-organizing societies. For example, modern societies lack the following characteristic that would emerge in a society in which the pursuit of self-interest would always tend to advance the common good:

(a) Corporations and other businesses that have no concern at all for the environment or for ethical behaviour would none-the-less act ethically and protect the environment.
(b) Politicians, political parties and governmental bodies would not have to be courageous, of good character or have political will in order to motivate them to act and govern in the interests of the society as a whole. Instead, the pursuit of societal interests would be the easy, self-interested thing for them to do.
(c) All members of the society would spend their lives helping others, even though they might not care about anyone else.
(d) Actions that contribute to the initiation of war within a self-organizing global society would be against the interest of corporations, political groups, and all other members of the society. In fact, it would be as much against their interests as war is against the interests of those whose lives are destroyed by it.

(e) Just as invisible hand processes currently ensure that some market processes serve societal goals, invisible hand processes would do the same for all the systems that establish governance and that undertake the other functions currently performed by governments.

(f) The society would be self-repairing and self-correcting in relation to any damage to its effective functioning.

(g) All power would be used in the interests of the society as a whole, whether it is possessed by individuals, corporations, governments or political groups.

(h) The development of better programs to help the disadvantaged fulfil their potential to contribute to society could be even more profitable and attractive to investors than developing a popular new breakfast cereal.

This brings us back to the key questions posed and addressed by this paper: is it possible to constrain a human society so that it is self-organizing in the sense discussed here? And if so, how?

In answering these questions, it is very useful to draw on the growing body of knowledge about how self-organizing societies of living organisms have emerged many times during the evolution of life on Earth [5, 6]. For example, societies of self-producing molecular processes emerged to produce simple cells; societies of simple cells produced eukaryote cells; societies of these cells emerged to produce multi-cellular organisms; and societies of these emerged to produce animal societies. Human evolution displays a similar pattern in which societies of increasing scale have emerged progressively: e.g. from kin groups to bands, to tribes, to city states, to nation states etc. In each of these instances, the societies which emerge at each step are formed of entities that previously lived independently and competed with each other. These entities are themselves smaller-scale societies that emerged previously in evolution. Furthermore, as we will see in more detail, all these societies that emerged in past evolution are self-organizing societies in the sense discussed here, with the exception of recent human societies.

For example, consider a multi-cellular organism such as yourself. You are a highly complex society comprised of trillions of cells. You are constituted by many more cells than there are humans on the planet. None of your individual cells knows anything of you, nor do any of them have any concern for your goals or interests. Each of your cells pursues its own immediate interests. It does only what cells do: acting on its cellular impulses, needs and volitions. Yet your cells produce you in all your complexity: your walking, talking, thinking and other functions. And as your cells adapt to their local circumstances in accordance with their immediate interests, they produce your adaptations and your achievements. How does this happen? We will use an understanding of how this has emerged to understand the conditions that must be met if human societies are also to be self-organizing.

**Section 2** of the paper sketches a general model of the emergence of societies/organizations that are self-organizing in the sense developed here. In developing the model, Section 2 draws on the growing body of knowledge about how self-organizing societies of living organisms have emerged many times during the evolution of life on Earth. Section 2 goes on to use the model to identify how a society needs to be constrained if it is to be self-organizing, and to understand the processes that have driven the emergence of such societies during past evolution.

**Section 3** draws on this general model to identify how self-organizing human societies can be developed, with particular focus on the establishment and adaptation of requisite institutions.

# 2. The Evolution and Emergence of Self-Organizing Societies of Organisms

### 2.1 Consequence-capture

This section uses an agent-based approach to develop a general model of the evolution of self-organizing societies. I begin by sketching a simplified model and then proceed to show how this simple model can be adapted to account for key features of the emergence of self-organizing societies at any level of organization.

Consider a population of agents that compete with each other to survive and persist. Agents can be self-producing molecular processes, proto-cells, prokaryote cells, eukaryote cells, multi-cellular organisms, animal societies, human tribes, corporations, nation states etc. Agents have a capacity to discover adaptations that can enhance their competitiveness. The nature of the mechanisms that adapt agents are not restricted—they can include processes as disparate as gene-based natural selection and psychological mechanisms. Ultimately agents compete to survive, but proximately they may compete to acquire psychological benefits, to increase profits or to otherwise increase utility.

Importantly, agents have the potential to interact with each other in ways that can impact on their competitiveness. Here we will explore the circumstances in which patterns of interaction between some agents within a population may emerge and be reproduced through time. Such a self-producing, self-maintaining pattern of entities and their interactions (relationships) would constitute an emergent organization within the larger population.

Actions taken by an agent that impact on other agents will not be adaptive for the agent unless the benefits it captures exceed the costs that it incurs. Actions that fail to produce net benefits to the agent will therefore not persist or be reproduced, and will not be part of any emergent organization. This is the case no matter how beneficial an action may be for the survival of the organization as a whole.

However, when an action benefits others and when the agent initiating the action captures sufficient of those benefits to outweigh the costs of the action, the competitiveness of the agent and of the other beneficiaries can both be enhanced. As a result, the relationships constituted by such actions will tend to be reproduced within the population. A self-producing/self-sustaining proto-organization will tend to emerge in a population whenever some agents capture sufficient of the benefits of their actions that also benefit other agents.

In summary, the general condition for the emergence and persistence of a proto-organization is ‘consequence-capture’—i.e. that agents capture enough of the benefits and harms of the impacts of their actions on others and on the organization as a whole to sustain them at an optimal level [6-11]. If this condition is met continually within an organization, it will constitute a self-organizing organization in the sense developed here: any action which an agent discovers that contributes net benefits to the organization as a whole will be adaptive for the agent and will be reproduced as part of the organization. The adaptive interests of the individual members of the organization will be aligned with the global adaptive interests of the organization. Agents that pursue their own interests successfully will therefore also be pursing the organization’s global interests. Agents will tend to adapt (and where applicable, evolve) in ways that improve the functionality of the organization as a whole. Through the adaptation of the agents that constitute it, the organization as a whole will be able to explore the space of cooperative organization, restricted only by the adaptability and evolvability of its agents. As we will see in greater detail, if this condition is not met the possibility space that can be explored by the proto-organization will be seriously limited, and the proto-organization will not be capable of open-ended evolution. Many forms of organization that contain highly beneficial cooperation will not be realizable.

It is worth emphasising here that these conclusions apply to all forms of self-organization involving living entities. These include the emergence of larger-scale order, patterns, structure and other forms of coordinated, global activity: if the emergent phenomena rely upon agents acting in ways that incur adaptive costs to them, the emergent pattern will not persist and be produced through time unless the agents capture sufficient benefits to outweigh those costs.

It should be noted that throughout this discussion, references to capturing benefits and harms that flow from the actions of agents refers to the benefits and harms that arise at all relevant levels and scales. It may be useful in some circumstances to consider the benefits (and harms) that are captured as comprising both a ‘between-group’ component and a ‘within group’ component (e.g. see [12]).

But how difficult is it for comprehensive consequence-capture to be achieved? And therefore how easily do organizations that are self-organizing in the sense developed here emerge in populations of living organisms? If this emergence was straightforward, we would expect it to occur frequently in populations because such organizations would be able to exploit the significant adaptive benefits of synergy and other forms of cooperation. There is widespread agreement about the potential benefits of coordination, division of labour, collective action and other forms of cooperation [13-15, 6]. In principle, these potential benefits apply in relation to organizations of all kinds of organisms, including multi-species assemblages. As a consequence, wherever such organizations emerge and persist, we would expect them to be strongly favoured by selection. This expectation is consistent with the evidence: whenever complex organizations have emerged in the history of life, they tend to be extremely successful [16, 5, 6-11].

However, even though such organizations are successful once they emerge, they emerge only infrequently. It is the exception rather than the rule. In the history of life, it is extremely rare for a population of organisms to give rise to complex cooperative organization, despite its adaptive benefits.

What are the reasons for this apparent impediment to the evolution of complex organizations? Why does comprehensive consequence-capture emerge only rarely? There is much research that helps answer this question. Due in part to interest in harnessing cooperation for human purposes, an extremely large body of research has focused on identifying the conditions that must be met if cooperation is to emerge amongst living processes [for overviews of particular parts of this literature see 17-21]. This research confirms that complex cooperative organization will emerge and persist only in limited circumstances [e.g. see 22-25].

I will draw upon this research to develop an understanding of impediments to the emergence of complex organization and to the achievement of consequence-capture. I will then build upon this research to examine how this impediment can be overcome in some circumstances through the emergence of special forms of organization that enable comprehensive consequence-capture and therefore enable the emergence of self-organizing societies.

### 2.2 How consequence-capture can be organized

This large body of research has identified a number of forms of organization that enable consequence-capture to some degree, at least in limited circumstances. **Figure 1** diagrams three forms of organizational architectures that enable some consequence-capture under some conditions. These architectures can account for the overwhelming majority of circumstances in which simple forms of cooperative organization arise in populations, in theory and in nature. The architectures are applicable irrespective of the nature of the particular agents that comprise the population and apply at all levels of organization.

**Figure 1. Architectures that allow limited consequence-capture**

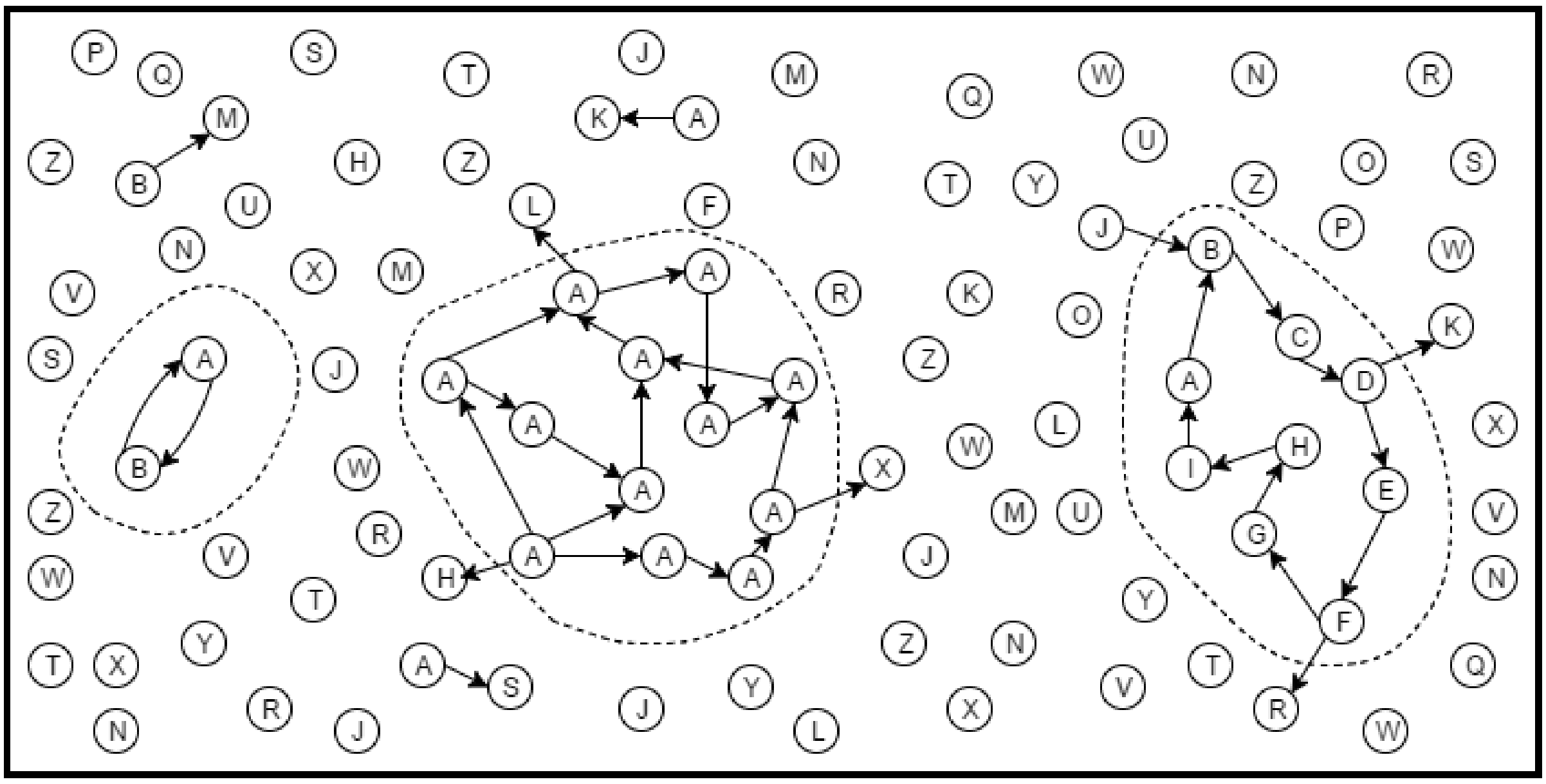

**Figure 1** depicts each of these three architectures within a larger population of agents that are not organized. Each kind of agent is represented by a circle containing a letter. Each organizational architecture is enclosed by a dotted line. Arrows represent the flow of benefits in the relationships between agents within an architecture.

Considering each organizational architecture depicted in **Figure 1** in turn from the left:

(a) In the first, each kind of agent captures some of the benefits it provides to the other when the other reciprocates. Note that the Figure also depicts a cooperative agent that provide benefits to a free-riding agent that fails to reciprocate. Specific examples of this architecture at various levels of organization include:

1) Reciprocal altruism e.g. where two humans provide food to each other when the other has insufficient for its needs [26, 27].

2) Mutualism e.g. where individuals from two different species each benefit from an association with the other, such as bees that feed from the flowers they pollinate [28].

3) Mutual catalysis e.g. between two different kinds of RNA molecule that each catalyse the formation of the other [29].

(b) In the second, agents that are predisposed to provide benefits to other agents capture some of the benefits of doing so when those other agents disproportionately share the same cooperative pre-disposition (each of the agents in the architecture are labelled ‘A’ to indicate they each share the same pre-disposition). Note that not all agents that share the pre-disposition receive benefits, and some provide benefits to free-riding individuals that do not share the pre-disposition. Examples of this architecture include:

1) Kin selection in which co-operators associate disproportionately with other agents who are similarly pre-disposed because they are related and share similar genes e.g. in some species of birds, individuals might feed the offspring of relatives [30, 31].

2) The cultural equivalent to kin selection in which the shared pre-disposition to cooperate results from shared norms or customs (rather than shared genes) e.g. some human ethnic groups share a cultural pre-disposition to preferentially do business with other members of their ethnic group [6].

3) Population structures (including separation into groups) that cause co-operators to associate disproportionately with other co-operators, and where groups with higher levels of cooperation may outcompete other groups [12, 33].

4) Selective cooperation in which, for example, human individuals are pre-disposed to cooperate only with others who have a reputation for cooperating (and may be predisposed to punish individuals that cheat or otherwise free ride) [26-27, 33-34].

(c) In the third, every agent that provides benefits to other members of the organization captures some of these because it is provided in turn with benefits by at least one other member of the organization. Note that some members of the organization provide benefits to free-riding individuals that do not contribute to the organization in return. Furthermore, one agent (J) contributes to the organization without being supported by it. Examples of this architecture at various levels of organization include:

1) A self-producing autocatalytic set of proteins in which each protein (i) catalyses the formation of at least one other member of the set, and (ii) is in turn catalysed by at least one other member of the set [35, 29].

2) An RNA hypercycle which comprises a self-producing family of RNA molecules that each (i) catalyses the formation of an enzyme which in turn catalyses the formation of another member of the RNA family, and (ii) has its own replication catalysed by an enzyme produced by another member of the RNA family [36].

3) Indirect reciprocal altruism e.g. a band of humans in which each member performs favours for others, but a favour may be repaid by members who were not recipients of the original favour [27].

4) An ‘autocatalytic cycle’ in an ecosystem comprising for example, a predator, a herbivore and grasses. The predator benefits by feeding on the herbivore, it benefits by feeding on the grasses, and the grasses benefit from the recycling of nutrients produced by deaths of the predator and herbivore and by their faeces [37].

This classification of architectures can be further compressed by noting that the first architecture is a special case of the third.

The reasons why these forms of organization are very limited in their capacity to realize the enormous adaptive advantages of complex cooperative organization are well known. In particular:

(a) The capacity of the organizations to produce themselves through time is highly susceptible to undermining by free riders. Free riders are agents that take some of the adaptive benefits produced by the organization but who contribute nothing (or insufficient) in return (e.g. cheats, thieves and defectors). Because free riders do not bear the costs of contributing appropriately to the organization, they can out-compete those that do. Furthermore, their appropriation of collective benefits prevents members of the organization from capturing sufficient of the benefits of their contributions. As a consequence, agents that would contribute significantly to the success of the organization are unable to persist in the organization. In these circumstances, the emergence of more complex forms of organization is not possible, despite their enormous potential adaptive advantages [24-25, 38].

(b) These organizational architectures do not contain any features which guarantee that agents will capture *sufficient* benefits to sustain *optimal* contributions to the organization. The architectures ensure that some agents will capture *some* benefits. But the architectures do not necessarily *match* the level of benefits to the costs of the agent's contributions. If complex cooperative organizations are to emerge, agents need to capture enough benefits to sustain their contributions at an optimal level, taking into account the costs of doing so [38, 6-9].

Due to these factors, the simple architectures that are generally invoked to explain the emergence of cooperative organization are unable to produce organizations or societies that are fully self-organizing in the sense developed here.

Because agents within these simple architectures are often unable to capture sufficient of the benefits of their actions (or harms in the case of free riders), the organizational possibility space that they are able to explore is seriously limited. There are many forms of organization, especially complex forms, that cannot be realized even though they include cooperative relationships that are clearly advantageous. Selection operating at the level of the individual agent or at the level of the organization as a whole is unable to overcome this limitation. If a form of organization cannot be reproduced through time, selection cannot call it into existence. Self-organization of the relevant variation must come first, or selection is irrelevant.

This is an observation that has significant consequences for understanding the evolution of living processes [6-7, 11]. But it is often not grasped fully. In evolutionary theorizing it is often accepted implicitly that the key challenge in any attempt to account for complex adaptation is to identify the selection pressures that have shaped and tuned it. This presumption works well when considering the evolution of organisms that are self-organizing, and are therefore capable of producing a more or less unlimited range of structures and functions on which selection can operate.

But this approach is completely inadequate for explaining the initial emergence within populations of the complex organizations that eventually became organisms in their own right (e.g. the emergence of the cooperative organizations of simple cells that became eukaryote cells, the organizations of eukaryote cells that became multicellular organisms, and the organizations of organisms that became animal societies). The assumption that the relevant variation on which selection can act will eventually arise and persist, fails to hold. Here the key challenge is instead to understand how complex, self-organizing organizations emerged out of earlier organizations that were incapable of sustaining complex forms of organization i.e. that were unable to produce complex variants on which selection could act. Demonstrating plausible selection pressures that would favour more complex forms of organization does not advance the explanation very far if these complex forms of organization are incapable of being realized by the simpler organizational architectures that precede them. For these reasons, the origin of life and other major evolutionary transitions (in which new, complex, larger-scale entities emerge from cooperatives of smaller scale entities) cannot be accounted for by selection operating solely on the kinds of architectures sketched in Figure 1. For example, a number of researchers have set out to show how evolutionary transitions can arise in a straightforward fashion out of the evolution of autocatalytic sets of macro-molecules and their architectural analogues at higher levels of organization (i.e. the kinds of architectures on the far right in Figure 1). However, these attempts have been unable to demonstrate how these autocatalytic forms of organization can evolve complex cooperative arrangements (for various attempts see 29, 37, and 40-41. For a critique see 42).

### 2.3 Architectures that provide for comprehensive consequence-capture

Are there more complex architectures that can enable full consequence-capture and therefore overcome the limitations of the simpler forms of organization considered in the previous section?

Stewart (6-11) demonstrates that the addition of a 'manager' (a system of evolvable constraints) to these simple architectures can achieve full consequence-capture. A manager is an agent or combination of agents that has the power to apply constraints across the simpler organization. These constraints can restrain or punish free riders that would otherwise undermine the organization. Constraints applied by the manager can also control the allocation of resources and other benefits within the organization. This enables the manager to ensure that agents that contribute to the success of the organization as a whole are supported optimally. By constraining free riding and supporting pro-social agents, the manager can ensure agents capture the consequences of their actions. Agents will therefore tend to adapt and evolve in ways that support the functionality of the organization as a whole.

A manager's capacity to constrain (and therefore control) the organization also enables it to harvest sufficient of the benefits produced by its management to support its own reproduction through time. This means that a manager will tend to capture benefits (and harms) produced by its management. As a result, consequence-capture will also tend to apply to the manager. It will therefore be in the manager's interests to manage the organization in ways that increase the effectiveness of the organization as a whole. The manager can best achieve this by ensuring that all other agents capture the impacts of their actions on the organization. The end result is that consequence-capture will tend to apply to all member of the organization, including the manager. This aligns the interests of the manager and all other agents with the interests of the organization as a whole. It produces a self-organizing organization in which any adaptations arising within the manager or other agents that enhance the success of the organization will tend to be supported and reproduced.

In summary, simple, unmanaged cooperative organizations tend to emerge at every level of organization, but the organizational possibility space they can explore is seriously limited. However, appropriate management (systems of evolvable constraints) can correct the limitations in these organizations, massively expanding the organizational possibility space that can be explored and enabling the emergence of complex cooperative organization, including major evolutionary transitions.

**Figure 2. The architecture of a managed organization**

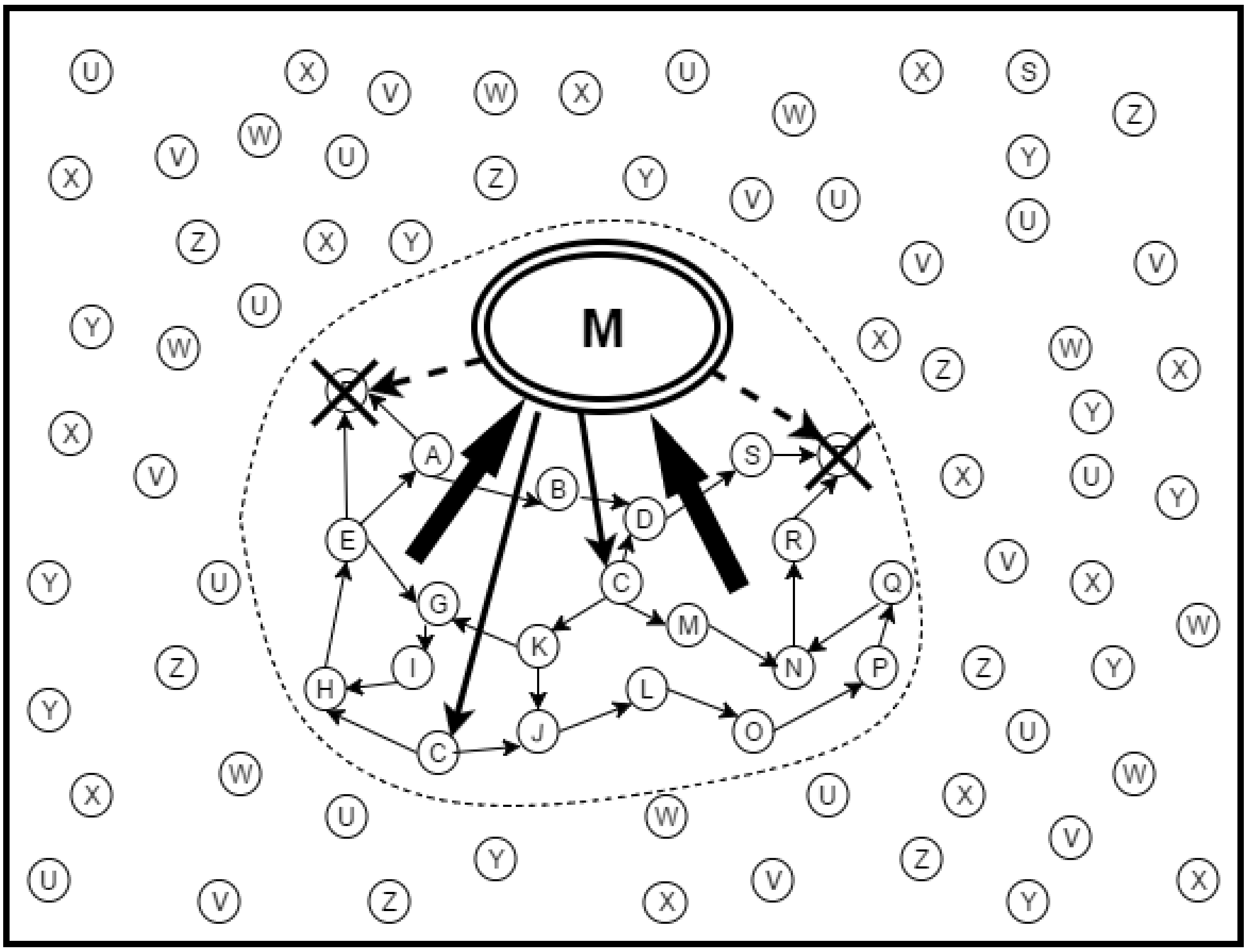

**Figure 2** sketches the architecture of a managed organization that is situated in a larger population of agents that are not organized. Each kind of agent is represented by a circle containing a letter. **M** is the powerful, evolvable manager. The normal arrows represent the flow of benefits between agents. The two dashed, bolded arrows that originate from the manager each represent the suppression of a free rider by the manager (each free rider takes benefits from other members of the organization but contributes nothing in return). Each suppressed free rider is marked with an **X**. The two bolded arrows (un-dashed) that originate from the manager each represent the provision by the manager of benefits to an agent that contributes to the organization but receives no benefits from other members of the organization (these agents would not persist in the organization if they were not supported by the manager, despite their positive contributions to the organization). The two thick black arrows pointing towards the manager represent the appropriation of benefits from the organization by the manager.

Examples of managed organizations at various levels of organization include:

(a) **RNA (and eventually DNA) management of proto-cells**. In early cells RNA management used its catalytic capacity to: (i) suppress side reactions and other free-riding metabolic processes that utilize the cell's resources without contributing to the functionality of the cell; (ii) support processes that contribute positively to the functioning of the cell but would not otherwise be reproduced within the cell; and (iii) support processes that assist management's own maintenance and reproduction. The threshold between chemistry and life was passed when RNA management took over proto-metabolisms and massively expanded the organizational possibility space that they could explore.

(b) **Management of a modern corporation by an effective CEO**. The CEO uses her or his power to manage the corporation by: (i) punishing and firing staff who free ride by failing to contribute sufficiently to the success of the corporation; and (ii) resourcing and incentivizing staff to the extent that the staff contribute to the effective functioning of the organization.

(c) **Governmental management of a nation state**. The government uses its power to manage the state by, for example: (i) punishing any free-riding citizens who don't pay appropriate taxes; (ii) using some of the taxes it raises to fund an army for the defence of the nation and to fund other activities that benefit the nation as a whole; and (iii) using some taxes to support the functioning and maintenance of the government itself.

In these three examples, the manager is external to the agents that are being managed. But Stewart also shows how management can also be instantiated by evolvable systems of constraints that are *internal* to agents, as well as by combinations of internal and external management. Internal constraints can control agents by pre-disposing them to behave in particular ways. These constraints can control an organization if they are reproduced within each of the agents that constitute the organization. Because such a set of constraints (collectively constituting the manager) can pre-dispose any member of the organization to act in particular ways, it is able to reach across and manage the entire organization. Because the manager is reproduced in each member of the organization, it will capture all the benefits (and harms) experienced by each member and by the organization as a whole. The manager will therefore be subject to consequence-capture, potentially meeting the key requirement for a self-organizing society. This consequence-capture will tend to align the adaptive interests of the manager with those of the organization as a whole.

**Figure 3** depicts the architecture of an organization managed by 'distributed internal management':

**Figure 3. Architecture of internal distributed management**

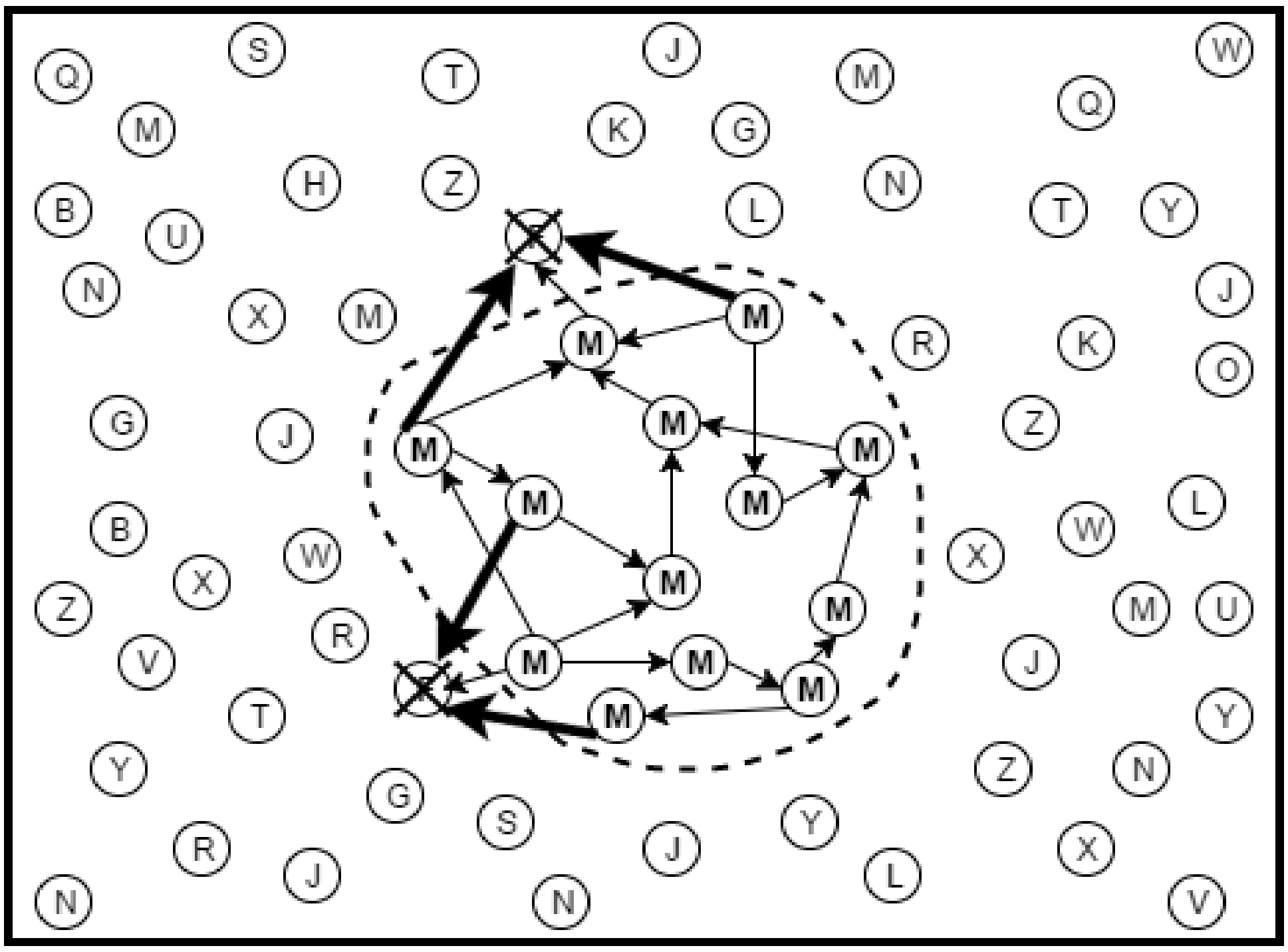

**Figure 3** sketches the architecture of an organization controlled by internal distributed management. Each kind of agent is represented by a circle containing a letter. Agents that include the bundle of predispositions that constitute the manager are marked with a bolded **M**. The managed organization is encircled by a dotted line. Normal arrows represent the flow of benefits between agents. All members of the organization contain the manager and cooperate with other members. There are two free riders depicted by circles marked F. The bolded arrows represent the suppression of free riders by a number of agents that contain the manager. Each suppressed free rider is marked with an **X**.

Examples of organizations managed by distributed internal management include:

(a) **Management of a proto-multicellular organism by distributed internal genetic elements**: An organization of cells can be managed by genes that are reproduced in each of the cells that constitute the organization. Because this genetic manager is included in each cell within the organization, it can pre-dispose cells to punish and suppress free riders such as mutants or invading cells that do not contain the manager. It can also pre-dispose cells to provide optimal resources to cells that contribute to the functionality of the organization as a whole.

(b) **Management of a human tribe by distributed internal norms and other cultural pre-dispositions**: A tribal society can be managed by norms and customs that are reproduced (internalized by socialization) in each member of the tribe. Such a manager can pre-dispose members of the tribe to punish or exclude any free-riding individuals who have failed to internalize norms that would otherwise have predisposed them to contribute to the success of the tribe. The manager can also pre-dispose members of the tribe to provide optimal support for members who, for example, specialize in providing useful services for the tribe, such as weapon making. This management by cultural constraints is likely to co-evolve with genetic management that comprises pro-social genetic pre-dispositions that are also distributed across the members of the tribe [6].

Even though distributed internal management arises 'bottom up' without any centralized external control, it operates in the same way as external management, and has the same effect on agents. It is equally constraining and controlling. However, where distributed internal management operates, it is not possible to perceive its management directly. This can lead to the misperception that the order it enables has emerged spontaneously without the involvement of any form of control at all. But agents that are managed by distributed internal management are not more free than those controlled by external management. In fact, they can be less free to innovate than individuals in externally-managed societies. This is because distributed internal management can only evolve and adapt when 'mutations' arise in the pre-dispositions that are hard-wired in all members of the society. But any 'mutant' individuals that are predisposed to act differently to other members of the society in ways that are socially significant will tend to be treated as free-riders and strongly suppressed [6]. Within organizations such as human tribal societies that were managed by distributed internal management, life for social innovators whose inventions contravened established norms tended to be nasty, brutish and short [61]. The difficulties facing social innovators also apply to some degree in human systems that are managed by a combination of distributed internal management and external management, such as those studied by Ostrom [43].

Whether management is instantiated by distributed internal management or external management or by a combination, consequence-capture will ensure that agents adapt and evolve in ways that enhance the functionality of the organization as a whole. But as mentioned in Section 1 above, this functionality itself will not necessarily be self-organizing. Whatever form of organization best serves the goals of the organization will arise and persist in a society which is self-organizing in the sense developed here. Top-down solutions will compete with bottom-up ones, mechanistic forms of organization will compete with more flexible and adaptable forms, and processing that is more or less distributed will compete with processing that is more or less centralized. The arrangements that win in one situation may change as circumstances and contexts change. If we use the organization of the human body as a guide to what kinds of outcomes might prevail, we can reasonably expect that a mix of all these forms of organization and processing will be called into existence in a complex society. Economic markets also point to a similar conclusion. They contain a wide mix of organizational types. In both the human body and in economic markets, no form of organization seems to be the most effective in all circumstances. For example, top-down processing often seems to prevail for systems that adapt organizations for what Beer [44] refers to as the outside/future—i.e. for interactions between the organization as a whole and its environment and/or in relation to future events. In these circumstances' most agents in the organization will not have the information they would need if they were to adapt directly for the outside/future. They will therefore need to have this information fed to them through some form of top-down processing. The way in which the brain functions in relation to the human body is a clear example. It is worth emphasizing that top-down processing does not necessitate rigid, prescriptive organization. Top-down control can be organized by enabling rather than prescriptive constraints, particularly where the members of the organization are sufficiently evolvable and adaptable in their own right [11]. As a further example of the contingent nature of organizational effectiveness, even flexibility and adaptability does not appear to be optimal in many circumstances. For example, where there is certainty and predictability, forms of organization that are relatively rigid and mechanistic can perform more effectively and efficiently than forms that waste resources on unnecessary exploratory activities, provided some residual adaptive capacity is retained.

## 2.4 Institutions and management

The systems of evolvable constraints that constitute management are generally termed 'institutions' in human societies. This terminology is consistent with North's observation that:

> Institutions are the rules of the game in society or, more formally, are the humanly devised constraints that shape human interaction. [45]

These constraints can arise from governance or other external sources, or from internal genetic and/or culturally-inculcated predispositions such as norms, or from combinations of both. The key characteristic of constraints in this context is that they can influence without being influenced in return [46]. If they could be influenced in return, they would not be able to control the behaviour of individuals and the distribution of resources within a society. This is because free riders and other individuals would be able to escape their influence. This capacity to constrain enables institutions to modify the interests of members of the society and to align them with those of the society as a whole.

In a self-organizing society, the constraining-power of institutions is used to implement consequence-capture, including in relation to the processes that establish and adapt institutions. Effective institutions in a self-organizing society ensure that when entities within the society (including individuals, corporations and governments) pursue their immediate self-interest, they will advance the goals of the society. As a consequence, it will be advantageous to entities to act in ways that advantage the society. In other words, in a self-organizing society institutions *enable* the self-organization of whatever patterns of behaviour serve the interests of the society as a whole. In short, institutions enable 'the good'. Importantly, this includes enabling the self-organization of whatever institutions are needed to further the interests of the society through the implementation of consequence-capture. A hierarchy of institutions will self-organize, ensuring that entities involved in the establishment and adaptation of institutions act in ways that serve societal interests. As a result, in a self-organizing society new institutions will tend to arise wherever consequence-capture does not apply—e.g. where there are conflicts of interest and/or interests that are not aligned with those of the society. As we shall discuss in the next section, when we evaluate institutions in current human societies against these standards, we find that many are highly ineffective. Institutions in modern societies often serve interests other than those of the society as a whole.

# 3. How can Self-Organizing Human Societies Be Organized?

The model sketched in Section 2 enables us to see broadly how human societies must be constrained if they are to be self-organizing. Management/institutions (i.e. systems of evolvable constraints) are needed to ensure that all agents in the society (including individuals, associations, businesses, corporations, politicians, political parties, institutions and governments) capture sufficient of the benefits (and harms) of their impacts on the goals of the society. Such management/institutions will align the interests of agents with those of the society as a whole. Agents that pursue their own immediate interests will therefore act in ways that also provide global social benefits, whether or not agents have any intention of doing so. Agents will adapt and evolve in ways that provide functionality for the society as a whole. Such a society will be self-organizing in the sense developed here.

What are the more detailed implications of this broad framework for how modern human societies might be re-organized so that they are self-organizing in the sense used here? In order to address this question, it is useful to begin by developing an understanding of the systems of management/institutions that have emerged during the evolution of human societies, and by identifying where they fall short of producing comprehensive consequence-capture.

### 3.1 The rise of externally-managed human societies

The earliest human societies (bands and simple tribes) were managed almost exclusively by systems of distributed internal constraints [6]. These were in the form of genetic and cultural constraints that pre-disposed individuals to behave pro-socially. Typically, this involved pre-disposing individuals to act cooperatively toward others who were also constrained by the same internal distributed manager (i.e. by the same bundle of pre-dispositions), and to punish or exclude others who were not (e.g. cheats, thieves and other free riders). For example, individuals in a tribe might have been socialized to internalize norms that pre-disposed them to reciprocate in exchange relations within the tribe and to punish individuals that cheated in exchanges [33].

External management emerged fully in human evolution only with the formation of larger societies such as agricultural communities, city states and kingdoms. Distributed internal management alone was not effective as a coordinator of behaviour across these societies. At least in part this was because of the unreliability of socialization as a mechanism for reproducing norms and other pre-dispositions across large numbers of people, often from different cultural backgrounds. Furthermore, as mentioned above, distributed internal management is less evolvable than external management in these circumstances [6]. The powerful rulers of these societies could manage more effectively, applying and adapting constraints in the form of governance that, for example, punished free riders and raised taxes to fund armies to defend the society and sometimes to plunder other societies.

Nevertheless, pro-social genetic and cultural pre-dispositions continued to play an important role in constituting institutions which managed interactions at the local level within these early externally-managed societies. They continue to do so in modern societies, particularly within sub-cultures. Furthermore, the rise of large empires and nations was greatly assisted by the emergence of 'world religions' that were capable of entrenching some pro-social norms across groups with different cultural backgrounds. These religions typically attempt to inculcate followers with broad injunctions that facilitate fairness, honesty and trust in economic and other social relations—e.g. 'to do unto others as you would have them do unto you' [47].

However, the significance of the systems of distributed internal constraints that were entrenched by religion has diminished somewhat in modern western societies with the rise of rationalism. The influence of religion was increasingly undermined as rationalism began to become more widespread with the emergence of the European Enlightenment. As a consequence, modern human societies have had to rely increasingly on external management in the form of governance to ensure that entities within it capture the benefits and harms of their impacts. These systems of governance include laws and associated enforcement processes, the raising of taxation, and the use of taxes to fund goods and services that are not provided by economic markets. The role of some of these institutions is particularly important because they enable the emergence of the dynamic webs of exchange relations that constitute economic markets. When it is effective, this governance enables exchange relations to flourish by punishing theft and cheating, including by enforcing contracts. In modern human societies the governance needed to underpin economic markets includes elaborate systems of laws, policing functions, courts and penalties. Without this complex and adaptable external management, modern large-scale economic markets would not emerge [48].

### 3.2 The lack of alignment between the interests of management and the interests of the society

In summary, external management predominates in modern human societies, but our societies still include diverse systems of distributed internal management that continue to play important roles in constituting institutions. However, as we saw in Section 1, modern human societies are not self-organizing in the sense developed here. In the context of the model sketched in Section 2, we can now see that this is because the systems of management in modern societies fail to fully align the interests of entities with those of the society. In order to understand why this is so and what could be done to change it, it is useful to look again briefly at the self-organizing organizations that have emerged during past evolution at lower levels of organization. These non-human organizations exhibit comprehensive consequence-capture and alignment of interests. We can explore why they succeed where modern human societies fail. Importantly we will find that this exploration points to how modern human societies could be re-organized and constrained so that they become fully self-organizing.

Stewart [6-11] shows that in non-human organizations, managers are likely to first emerge as powerful agents that plunder organizations of entities. Under favourable circumstances, these powerful agents may discover profitable ways to harvest an on-going stream of benefits from an organization, rather than exploit and plunder it only once. This begins the alignment of interests between the proto-manager and the organization it manages. The alignment is strengthened as managers discover how to constrain the organization in ways that increase the productivity and survivability of the organization (e.g. by promoting useful cooperation). This alignment will be further strengthened once the manager is obligatorily tied to one organization and cannot move between them (i.e. so it is no longer able to avoid the consequences of any over-exploitation of the organization). Full alignment will be achieved when competition between managed organizations within a population ensures that any action taken by the manager that is contrary to the interests of the organization will be strongly selected against i.e. poorly-managed organizations will be out-competed and eliminated from the population. In these circumstances, the only way the manager will be able to advance its own interests is by applying constraints which advance the interests of the organization as a whole.

But full alignment is obviously not a feature of modern human societies. Individuals and organizations that constitute management/institutions in human societies can often advance their own interests to some degree at the expense of the society. The most significant reason for this is the absence of strong competition and selection at the level of human societies. This is a consequence of the absence of a large population of human societies that are in strong competition with each other for limited resources. There are fewer than 200 sovereign countries on the planet today. If circumstances producing strong competition did exist, societies whose managers were relatively ineffective would be out-competed and quickly eliminated from the population. Ineffective managers include those that exploit their society by appropriating more of the resources of the society for themselves and their supporters than is optimal.

In modern human societies, weak competition and selection at the level of societies can allow governments and the interests they serve to get away with exploitative and ineffective management. As a result, powerful interest groups are able to co-opt the power of government to further their own interests at the expense of the society, including by distorting and manipulating the framework of institutional constraints that are essential if economic markets are to operate efficiently [49-53]. This weakness of competition and selection has progressively worsened in the most recent 10,000 years of human evolution as the scale of human societies has increased and the numbers of human societies has progressively decreased. It will peak if the trend continues and eventually leads to the emergence of a single human society on the scale of the planet.

But the solution is not to reduce the scale of human societies and to promote fierce competition between them. More effective and efficient ways to re-align the interests of government with those of the society have begun to emerge during the recent evolution of human societies [6]. These arrangements that have begun to emerge are internal to the society itself. They operate by constraining the agents, processes and systems that establish and adapt the governance and other institutions of the society. The effects of these 'meta-constraints' is to begin to achieve the same outcomes through *intra-societal* processes that strong competition and selection would achieve between societies—i.e. to implement intra-societal processes that begin to align the interests of governance with the interests of the society as a whole by ensuring that management captures the benefits and harms of the impacts of its actions on society.

An early step in the development of these intra-societal arrangements was the signing of the Magna Carta in Britain in 1215. For the first time, the king himself was made subject to the laws of the land. These laws constrained the king to act in ways that were beneficial to some degree to some of the other interests in the society (it was a small, first step). Representative democracy was a further major step in the emergence of internal constraints on governance. It ensured that the individuals who were given the primary power to establish and adapt governance were subjected to regular election by members of the society.

### 3.3 The limitations of democracy as a method for constraining management/governance

The democratic systems in place in the world today have gone some way towards constraining governments so as to align their interests with those of citizens. But they fall far short of doing so comprehensively. In part this is because citizens are not the only interests in society who are attempting to constrain governments to act in their interests. Other key players include individuals who have accumulated much more of the societies' resources and wealth than is necessary to sustain their positive contributions to the society (e.g. various kinds of free riders). These interests have powerful incentives to influence governments to allow them to maintain and expand their wealth and power. Furthermore, the royal road to riches in managed human societies has always been to co-opt the power of government to distort economic markets and other aspects of society in your favour. The wealthy and powerful are generally the best-positioned within society to take advantage of this principle. They have the incentive and the means to manipulate governments for their own ends [25-29]. This has long been recognized by economists with an interest in institutions. For example, Adam Smith wrote when talking of businessmen:

> The proposal of any new law or regulation of commerce which comes from this order, ought always to be listened to with great precaution, and ought never to be adopted till after having been long and carefully examined, not only with the most scrupulous, but with the most suspicious attention. It comes from an order of men, whose interest is never exactly the same with that of the public, who have generally an interest to deceive and even to oppress the public, and who accordingly have, upon many occasions, both deceived and oppressed it. [4]

And more recently, Nobel Prize winner Douglas North wrote:

> Institutions are not necessarily or even usually created to be socially efficient; rather they, or at least the formal rules, are created to serve the interests of those with the bargaining power to devise new rules. [45]

Just as the wealthy and powerful pursue their immediate self-interest in economic markets, they also purse their immediate self-interest to influence governments to establish institutions that serve their interests. But unlike in ideal economic markets, there is as yet no invisible hand to ensure that their pursuit of self-interest in their manipulation of governance will also serve the common good.

But why cannot citizens use their power at the ballot box to ensure governments only act in the interests of the majority of citizens? Firstly, this power is very weak. It is typically only exercised every three or four years. Furthermore, voters do not have the power to impose their choices separately in relation to each of the multitudes of initiatives that governments will undertake in their period of office. Instead, voters only have the power to choose one of a very small number of alternative packages of initiatives (often only two political parties have a realistic chance of gaining office). They have no option but to take the bad with the good, and may end up agreeing with only a small proportion of the package of proposed initiatives that they vote for. They cannot pick and choose initiatives out of each package.

It is worth imagining how unresponsive and ineffective economic markets would be if they operated under similar principles—i.e. if consumers had to purchase all their goods and services in advance for a four-year period through a collective decision that is able to choose just one of only two or three packages of goods and services. If economic markets were as restricted and limited in these ways as political markets, they would not have given rise to the enormous diversity of goods and services tailored to individual needs and wants that we see in modern societies. They would be just as incapable of aligning the interests of producers and consumers as are current democratic systems at aligning the interests of the governors with the interests of the governed.

This weakness in the capacity of citizens to constrain their governments in modern democracies leaves ample space for the wealthy and powerful and other interests to influence and manipulate governments for their own ends. Those who have grown wealthy and powerful under the existing organization of society are well-placed to manipulate government. They can use their wealth and power directly to corrupt the government. But in sophisticated modern societies they can use means that are far more subtle. Wealthy owners of mass media who share common interests with other powerful individuals use their control over public debate to influence voter's decisions. Mass media have a strong hand in determining what public issues will be treated seriously, what opinions will be repeatedly published and which will be taken as authoritative. They also largely control the extent to which the background, actions and motivations of politicians are critically scrutinized, and the extent to which their failings are brought to the attention of the public [49, 52].

The wealthy and powerful are also in a position to co-opt governments and the mass media to use war, hatred and fear to manipulate voters. It has long been recognized that many citizens forget their preference for a cohesive, caring and fair society when they believe that their way of life is under serious threat from enemies or other crises, whether internal or external [51, 52]. As Plato pointed out during the dawn of democracy in ancient Greece, citizens will overlook almost any failings of government if their society is at war, either against other states or against other races, ethnic groups or religions within the society [54]. A politician's dream has always been a war that can never end, such as an international war against terrorism.

The manipulation of citizens in modern democratic societies is greatly assisted by the existence of a significant proportion of citizens who have not yet reached the level of cognitive development that began to spread more widely with the European Enlightenment—i.e. broadly the formal operations level identified by Piaget [55]. Even in modern industrialized societies, only about 30 per cent develop this level of cognitive capacity. The remainder are at Piaget's concrete operations level at best [55-57]. As such, these citizens do not have an independent capacity to use abstract ideas and principles to judge and evaluate the motivations and effects of the actions of government. Unable to think abstractly for themselves, they have to rely on the thinking of 'reputable authorities'. And they do not have the capacity to assess for themselves which of these authorities are, in fact, reliable.

The manipulation and distortion of governmental activities is further assisted by the extreme centralization of control over governance in modern societies. The enormous power of government to intervene in and control all aspects of society is in the hands of typically a few hundred individuals for three or four years at a time. If sufficient of these are influenced and controlled, the society is controlled.

But none of this is to say that governments and voters are controlled and manipulated by some kind of organized conspiracy. This control and manipulation self-organizes out of the individual interests of the wealthy and powerful. Individuals simply act in accordance with their own perceived wider interests. Their wider interest is often as simple as maintaining the status quo of governance and policies that have enabled them to accumulate their wealth and power. There is no central body that plans and directs strategies designed to advance the collective interest of the wealthy and powerful. And because the collective action that emerges amongst the wealthy and powerful self-organizes from individual interests, it is also limited and undermined to some extent by the impediments to collective organization outlined in Section 2 [24]. For example, individuals are unlikely to invest significant resources in pursuing the collective interest where they capture insufficient of the benefits to cover their investments. And various forms of free riding can be expected to be rife (there is no honour amongst thieves). But nonetheless, in large part due to the centralization of government and the influence of mass media, the ease with which government and voters can be manipulated enables this flawed self-organization of collective power to be relatively effective.

### 3.4 New approaches to aligning the interests of management with those of the society

How could modern societies be re-organized and constrained to overcome these problems?

The general understandings developed here about how to produce a self-organizing society point to a way forward. These understandings suggest a method for diagnosing whether a society or any of the sub-systems or processes within the society will self-organize in the interests of the society. Where this diagnosis reveals that the sub-systems and processes will fail to self-organize appropriately, these understandings also suggest in general terms what must be done to rectify the situation. It identifies the actions that can be taken to re-organize the sub-systems and processes so that they henceforth self-organize 'the good', as the society defines it.

In relation to diagnosis, the fundamental investigation to be undertaken is this: do all relevant agents (including agents that are organizations of agents, such as corporations) capture sufficient of the benefits (and harms) of their actions to ensure that pro-social actions will be reproduced optimally in the society? If all relevant agents pursue their own immediate interests, will they act in the interests of the society as a whole? The investigation needs to identify all situations and circumstance in which the interests of any specific agents conflict with societal interests.

In any particular instance, this investigation must also be undertaken in relation to agents and processes that establish and adapt relevant governance and other institutions. This is likely to require an examination of the relevant government agencies and associated processes as well as politicians and political processes. Whether consequence-capture applies must be assessed for all agents and processes, including those involved in management.

Where such an investigation reveals that particular agents and processes do not capture sufficient of the relevant benefits and harms, systems of adaptable governance and other institutions need to be put in place to ensure that they do. This will involve embedding all relevant individuals, businesses and other agents in systems that provide them with a pattern of incentives, disincentives and other constraints which align their interests with societal goals. Incentives may include direct funding, opportunities to make profits, salaries, career advancement and psychological benefits. Disincentives may take the form of penalties, sanctions, loss of financial benefits and taxes (e.g. taxes on carbon emissions). Other constraints can include laws, regulations, rules and associated enforcement processes. In circumstances where local information is important, management will not be prescriptive about what agents should do. Instead it will set goals and allow agents to discover the best ways of achieving them, taking into account local circumstances [11].

Taken together, these systems of incentives, disincentives and other constraints would ensure that the social consequences of the actions of all relevant agents are fed back to them. This would ensure that the only way agents can satisfy their own immediate interests is to serve ‘the good’.

Of course, if this approach is to be fully self-organizing in the sense developed here, the agents and processes that put in place this new management (including the agents involved in undertaking the initial investigation which assesses whether new management is necessary), must also be governed and managed so that they are subject to consequence-capture, as do the processes that put them in place, and so on, with ultimate closure at the level of the constitution. A hierarchy of management is necessary, with each level constraining the one below and constrained by levels above.

In many situations, there will be uncertainty about the particular form of management that will be optimal and which should be implemented. For example, uncertainty might result from the complexity of the circumstances or lack of relevant information. Where this is the case, decisions about governance and management will generally be more effective if they are made by processes that are distributed and adaptable, rather than made centrally by an individual or agency [6, 11]. Effective decisions about management would not be made by individuals or by government agencies, but by processes constituted by many agents, possibly competing with each other to develop the best decisions. These distributed processes would need to be organized so that agents within them capture the benefits and harms of their actions i.e. they also would be self-organizing or ‘invisible hand’ processes.

However, an approach that sets out to develop a self-organizing society using this kind of diagnosis and re-organization on a piecemeal, ‘sub-system by sub-system’ basis would be expected to encounter strong resistance. This would come from any powerful interests that would be affected adversely by each initiative. If each step in such a piecemeal approach to reform were to be successful, it would have to begin by appropriately constraining the powerful interests that were threatened by it. Attempts to reform an entire society by such a step-by-step approach could also begin strategically by commencing with areas of governance where the impact on powerful interests was least. However, such a strategy would likely soon be detected by those interests, and opposed.

### 3.5 Alignment of interests through vertical markets

An alternative to a reformist, piecemeal approach would be to attempt to implement a self-organizing society in ‘one fell swoop’ (this might be possible for example, in times of crisis where possibility space is considerably expanded. For more detailed discussion about how a political movement might be developed that could take advantage of such circumstances to move towards a self-organizing society, see Stewart [58]).

How would such a society need to be organized? What would it look like as a functioning whole? Is it possible to envisage new systems for establishing governance and other institutions that would be capable of producing comprehensive consequence-capture so that the society would thereafter function and maintain itself as a self-organizing whole?

We can begin to answer these questions by first developing an understanding of the strengths and weakness of economic markets. In particular, an understanding of how markets can operate as self-organizing systems will enable us to see how the processes that establish institutions could also be structured so that they are self-organizing. Appropriately-managed economic markets are able to incentivize the production of an extraordinary diversity of goods and services that are able to satisfy a wide range of human needs and wants. Economic markets achieve this by utilizing webs of competitive exchange relations which enable agents to capture benefits from the provision of goods or services that satisfy the needs of others. Exchange relations are key to the self-organizing capacity of markets. They enable market participants to capture benefits by helping others who they may never meet. They enable comprehensive consequence-capture for the satisfaction of the needs and wants that economic markets can service efficiently. It is this characteristic of markets that has unleashed exponentially-increasing creativity, innovation and investment in recent centuries.

But there are many needs and wants that economic markets are unable to address. Economic markets provide consequence-capture for the satisfaction of only a subset of human preferences and desires. For example, the exchange relations that constitute markets can operate effectively only for goods and services whose benefits are discrete and can be enjoyed only by the purchaser. If this central condition is not met, free riders will be able to enjoy the benefits of a good without paying. This will prevent the producer from capturing sufficient of the benefits generated by such a ‘public good’. Other market failures can also undermine the capacity of economic markets to provide appropriate consequence-capture for some goods and services, including where there are natural monopolies, principal-agent problems, time-inconsistent preferences, externalities, information asymmetries, and so on. As a consequence, economic markets can fail to provide efficiently a range of goods and services that are necessary to satisfy particular human needs and wants, including: the laws and other institutions that enable effective economic markets to emerge in the first place (including the constraints that are needed to deal with externalities such as those that drive global warming); the goods and services needed by those who do not have sufficient purchasing power in the economic market; the infrastructure and constraints needed to underpin flourishing and satisfying communities; environmental protections; aspects of education and other initiatives designed to develop the full potential of all citizens to contribute to the society; defence; law creation and enforcement; initiatives to promote effective forms of internal distributed management within society such as trust and pro-social norms; solutions to collective action problems; the provision of services that tend to be local monopolies; other sets of constraints that can assist in ensuring consequence-capture; and many other functions currently performed by governments.

Taking this context into account, a key challenge that can be posed in relation to current human societies is this: how can societies be re-organized so that the satisfaction of *all* needs and desires (including collective needs) tends to be addressed optimally by self-organizing (invisible hand) processes? Stewart [7, 9, 6] shows how this can be achieved by what he terms a 'vertical market'. This is a web of competitive exchange relations that enable consequence-capture for 'products' that satisfy the needs and wants that are not able to be addressed by economic markets (i.e. by horizontal markets). The basic element of a vertical market is an exchange relation in which public goods, institutions and other forms of governance are 'purchased' by the *collectives* that benefit from them. Entrepreneurial 'producers' compete to offer 'products' for sale to *collectives*. Through the sale of these 'products', entrepreneurs capture social benefits produced by the public goods, institutions and other services they develop. Collectives decide whether to purchase a particular component of governance using for example, a voting process. If the decision is made to buy the good, the members of the collective are *required* to contribute appropriately to the purchase (overcoming the free-rider problem at the level of the collective). This compulsory contribution to a collective purchase is a counterpart of taxation in current societies.

Both vertical and economic markets are complex webs of exchange relations in which producers compete to satisfy the preferences of consumers. But it is worth emphasizing here two fundamental differences between vertical and horizontal markets:

(a) First, the products exchanged in vertical markets differ in an important respect from those exchanged in economic markets. A key feature of these products is that they usually provide benefits to collectives rather than to individuals, often unavoidably so. In contrast, goods and services that are exchanged effectively in a horizontal market are products whose benefits can be restricted to the specific agent that purchases them. Many of the goods and services currently provided by governments are examples of components of governance and public goods that cannot be purchased in economic markets. Numerous examples have been given above, and include the complex framework of regulations and other constraints (i.e. the institutions) that enable both horizontal and vertical systems of exchange relations to emerge and operate in the interests of society.

(b) Second, each exchange that occurs in a vertical market is between a 'producer' and the *collection of agents* that benefits from the component of governance that is being exchanged. In contrast, each exchange in an economic market is between a producer and *an individual agent* that solely enjoys the benefits of the good. This feature of vertical markets is of critical importance to their successful operation. If an effective system of exchange relations is to emerge, all of the agents who enjoy the benefits of an instance of a product must be involved in its purchase so that the producer can capture sufficient of the benefits provided by the good.

    To ensure this condition is met in vertical markets, purchasing decisions need to be made by the collection of agents who will enjoy the benefits of a particular component of governance that is on offer. An important part of the framework of vertical markets will be processes that identify the constituents of the relevant collective in any instance.

In cases where the benefits of a public good or other component of governance are enjoyed by everyone in a society, the collective that decides whether to purchase the good will comprise all citizens. But the collective will be smaller where the good benefits only a subset of the society—e. g. where the product is a training scheme designed to correct the under-provision of training in an industry that results from the inability of individual employers to capture sufficient of the benefits of training they provide to their employees. In this example, the relevant collective will be the employers in the industry. If the collective decides to buy the scheme, all employers will be taxed to fund it in proportion to the net benefits they receive. By way of a further example, where the product is governance that restrains environmental harms, the nature of the collective that benefits from the governance will depend on the scale and location of the environmental harms. Some products may also have negative impacts on other agents. In these cases, the framework that regulates the vertical market could require the producer to enter into an exchange relation with the collective that experiences the negative consequences in order to ameliorate the harmful impacts. The hierarchical systems of regulation and other constraints that provide the framework for the operation of a vertical market would themselves be established and adapted through the vertical market system.

Through vertical markets, collectives would be able to purchase goods and services that are directed at satisfying needs and wants that cannot be adequately addressed by economic markets. The *combination* of vertical and horizontal markets would enable comprehensive consequence-capture for those involved in producing goods and services designed to satisfy the full range of human wants and need. The result would be a society in which actions directed at satisfying *all* human needs would tend to self-organize. Vertical markets would replace the functions performed by centralized government. Because vertical markets are distributed rather than centralized, they would be far less susceptible to manipulation by powerful interests, enabling much closer alignment of interests between the ‘producers’ and ‘consumers’ of public goods and other elements of governance.

Other key features of a vertical market system include:

(a) Entrepreneurs would be able to capture benefits from their investment in the development and production of better institutions and other forms of governance. This is a critically important feature of vertical markets that distinguishes them from current democratic systems. Because entrepreneurs who develop and implement products for the vertical market would be able to capture these benefits, they would find it profitable to make large investments in research and development where these are necessary to realize the huge benefits that can flow from better institutions and governance. In relation to improvements in governance, this would unleash the same kind of exponentially-increasing innovation that we see in economic markets.

This capacity of vertical markets also distinguishes them sharply from proposals to introduce new forms of democracy which are more distributed, including proposals for e-democracy that take advantage of recent developments in information and communication technologies (e.g. see [59]). In these proposals, citizens would be able to vote directly on each of a wide range of policies and other aspects of governance, including proposals that can be developed by citizens themselves. These 'hyper-democratic' schemes contain a number of advantages over current forms of democracy. However, they also contain a serious impediment. They lack market processes that allocate appropriate levels of resources and investment to the development of innovative proposals for new public goods and other elements of governance. As we have seen, they share this deficiency with current forms of democracy. This impediment arises because any individual will only capture a small proportion of the potentially huge benefits that can flow to a society from more effective governance. Free riding will undermine collective efforts. As a consequence, individuals will not have the incentive to make the significant investments that may be needed to research and develop better forms of governance (a classic example of a collective action problem). Our current systems typically provide much greater rewards and resources for the development of a tastier breakfast cereal or ice cream than for the development of better market regulation.

Hyper-democratic schemes that do not contain self-organizing processes that allocate optimal levels of resources for the development of public goods and other forms of governance are seriously limited. They would be as ineffective as economic markets that were limited in a similar way. Imagine an economic market that did not allocate any resources to entrepreneurs for the development of innovative products and services. Instead it left it to consumers to develop new products and services. If economic markets were as limited as democratic systems in this regard they would never have produced the great diversity of goods and services that flood the modern marketplace. Proper resourcing of product development is the engine of innovation and creativity in both vertical and horizontal markets.

(b) Every unmet (or inefficiently met) need in a self-organizing society would be a profit opportunity in either a vertical or horizontal market. Because entrepreneurs could capture the benefits of their contributions, they would be incentivized to seek out and rectify any loophole in governance or deficiency in existing public goods. And as changing circumstances render particular elements of governance sub-optimal, entrepreneurs would profit from adapting governance to the new conditions. Corporations that found ways to game the system would create profit opportunities for entrepreneurs who devise ways to counter them. Entrepreneurs would be able to profit from identifying and correcting free-riding, uncaptured harms, inefficient governance, governance that fails to make effective use of emerging technological advances, agents that capture greater benefits than are appropriate given their contributions to society, the manipulation and distortion of governance and markets, and so on. As they respond to these opportunities, entrepreneurs would serve the interests of the members of the society even if they were motivated to pursue only their own immediate interests. Societies with effective vertical markets would therefore be self-organizing, self-repairing, self-improving, and self-adapting.

(c) Vertical markets would also be self-evolving. This is because the framework and regulation of a vertical market are also products that could be purchased in the vertical market. Everything in a vertical market would therefore be contestable and evolvable, including the institutions of the vertical market itself. Any valid criticism of any aspect of a vertical market would not be fatal to the vertical market. It would instead point to a profit opportunity associated with the correction of the deficiency.

As mentioned already, the regulation of horizontal economic markets would also be dealt with by the vertical market. Vertical markets would incentivize the development of regulations and other constraints that would prevent powerful interests from distorting and biasing economic markets in their favour. The full potential of economic markets to satisfy human needs within their limited domain could only be fully realized once the governance that manages them is as adaptable, innovative and creative as the products produced by economic markets.

The vertical market system would be a meta-market (a market in markets, both vertical and horizontal). It would 'decide' which particular needs would be addressed by products exchanged in vertical markets and which by horizontal markets, and how this might change as circumstances change.

(d) Individuals would not be overwhelmed by the demands of decision making in the vertical market. They would be free to choose which particular vertical market transactions they became involved in personally (if any). Where they choose not to be involved, they could delegate their vote to processes that utilize collective intelligence such as citizen juries or citizen deliberative councils, or delegate their vote to other citizens or even to representatives (e.g. see [60]). The vertical market could be expected to respond to the diverse needs and preferences of citizens by providing a wide array of options for decision making.

The important point is that citizens would be able to delegate their decision making to processes that are far more intelligent and knowledgeable than they are. This would be critical to the effectiveness of vertical markets. The ability of a vertical market to establish and adapt appropriate governance would depend in large part on the competence of the decisions made by or on behalf of citizens. The ability to delegate decision making to processes that are more intelligent would also help overcome one of the key deficiencies in current democratic systems: as mentioned earlier, in existing systems the wealthy and powerful can take advantage of and exploit the cognitive limitations of many voters.

(e) As it does in economic markets, competition between 'producers' in the vertical market would drive the alignment of interests between 'producers' and 'consumers' of governance 'products'. Competition would tend to ensure that 'producers' would benefit only to the extent that they could best satisfy the interests of the collectives that are the 'consumers' in a vertical market. Also as in economic markets, this competition would be intensified and creativity would be maximized by allowing open access to the vertical market: any individual, business, corporation or other organization would be able to assume the role of an entrepreneur who develops and offers new 'products' in the vertical marketplace.

(f) The combination of vertical and horizontal markets would tend to call into existence whatever arrangements would optimally satisfy the preferences of the members of the society, subject to resource and other constraints. Economic markets alone cannot go anywhere near achieving this.

(g) A vertical market would not eliminate governance and other forms of management. These are ineradicable features of self-evolving societies at all levels of organization. They are essential if citizens are to capture the benefits and harms produced by the impact of their actions on the goals of the society. However, the vertical market could be expected to constrain management so that it achieves its goals in ways that minimize restrictions on the freedom of citizens and other agents.

(h) The vertical market would 'decide' what kinds of organization are established within the society to serve the society's goals. As discussed earlier, self-organizing societies in general have no intrinsic preference for particular forms of organization. Whatever forms of organization will best serve the interests of the society in the prevailing circumstances can be expected to prevail. However, as the intelligence and evolvability of members of societies has increased during the evolution of life, there is a clear trend towards forms of organization that are less prescriptive and that better utilize the creativity of their members. This is evident in human evolution—the rise of market-based societies is due in part to their ability to enable a much higher proportion of citizens to engage in innovation and creativity than do feudal societies, for example.

Of course, it would not be necessary to implement a fully-fledged vertical market system all at once. Instead, a vertical market system could be introduced in a phased, step-by-step manner. If, for example, the starting point was a typical modern society with a system of representative democracy, a vertical market could be established initially to adapt only a small sub-set of governance, or even just a single aspect of governance (the burgeoning expansion of 'special district governments' in the United States that is outlined by Clarke [61] can be interpreted as a step in this direction). Additional systems of governance could be added to the vertical market progressively over time. Of course, as for any step-by-step approach, powerful interests could be expected to oppose and distort any moves to implement even a very limited vertical market that threatened their interests. A first step in any such approach would therefore need to involve constraining those interests appropriately.

### 3.6 The need for global management/institutions and a planetary self-organizing society

A vertical market system operating within a society has the potential to ensure consequence-capture *within* the society. However, it would not ensure consequence-capture for actions that have impacts outside the society, including interactions with other societies. More specifically, even if each nation on Earth were a self-organizing society with full internal consequence-capture, cooperation between nations would still be impeded. Thieving between states (i.e. war) and other forms of free riding would continue to undermine the ability of nations to cooperate together even when it would benefit them and their citizens. Nations would not capture the benefits or harms of the impact of their actions on other nations.

The absence of consequence-capture at the international level can produce serious problems, particularly where there are uncaptured environmental harms such as those caused by excessive carbon emissions. In the case of global warming, businesses and other agents do not currently capture in full the global harms caused by their release of greenhouse gasses into the atmosphere. As a consequence, global warming is self-organizing apace. But any nation that attempts to stop this by ensuring that its businesses and other agents capture these global harms will be disadvantaged relative to nations that do not. Such a nation will fail to capture the benefits of taking this action. In contrast, free-riding nations that fail to restrain the carbon emissions of their corporations and other businesses will be advantaged—their businesses will be able to out-compete those in other nations that are not free-riding [62].

For these reasons, a system of global management is essential if the citizens of the world and all other agents (including nations) are to capture sufficient of the benefits and harms of their impacts on others [6, 11]. An appropriate system of global governance and other global institutions is necessary to align the interests of individuals, organizations, businesses, corporations, politicians, governments and all other agents with the interests of global society. This would ensure that when agents within the global society purse their own immediate interests, they will also serve the goals of humanity as a whole. The result would be a self-organizing society on the scale of the planet.

The general principle here is that if ‘the good’ is to self-organize, consequence-capture must apply over all relevant scales of space and time.

# 4. Conclusion

We have seen that it is possible to organize a human society in such a way that actions which advance the goals of the society will tend to self-organize. In such a self-organizing society, individuals and other agents that seek only to pursue their immediate interests will also advance the societal interest. Importantly (and fortunately for humanity), a society does not have to comprise agents who want to serve the interests of the society in order for the society to produce ‘the good’.

We have learnt from an examination of self-organizing societies that have emerged previously during the evolution of life that the key to organizing such a society is consequence-capture. All members of the society must capture sufficient of the impacts of their actions on the society as a whole. This can be achieved by sets of evolvable constraints that I have termed ‘management’. In human societies these sets of constraints are commonly referred to as institutions. Management constrains the society to prevent free riding and to support contributions to the goals of the society. Unconstrained or inappropriately-constrained self-interest can destroy a society. Appropriately constrained self-interest will produce a society in which ‘the good’ tends to self-organize.

In the absence of consequence-capture the ‘possibility space’ for a society is seriously limited. There will be many actions and processes that are beneficial to the society that will not emerge and that will not be sustainable within the society. This is because agents will not capture sufficient of the benefits that would accrue from investing resources in these actions and processes. Furthermore, agents that do invest in these actions and processes will tend to be out-competed by free riders that do not.

These considerations point to an important principle for dealing with problems and challenges in human societies. It is not enough to devise a technical solution to any problem. The solution will not be implementable unless relevant agents capture sufficient of the benefits of doing so. Any attempt to solve a complex societal challenge should therefore always begin with the meta-question: will all relevant agents experience sufficient consequence-capture as they search for and implement an appropriate solution? If they will not, the problem is likely to be insoluble in practice. If they will, self-organization will tend to ensure that agents search possibility space for an optimal solution, implement the best that is found, and adapt the solution as circumstances change. Provided appropriate enabling institutions which ensure consequence-capture are in place, the relevant agents can often be left alone to solve the problem without further intervention (note that these considerations apply equally to strategies designed to establish a self-organizing society).

If a society is to be self-organizing, consequence-capture must also apply to agents that are involved in establishing and adapting management. This challenge is particularly important for modern human societies. In contrast to non-human societies at lower levels of organization, selection at the level of modern societies is often not strong enough to fully align the interests of management with those of the society. Human societies must therefore develop processes internal to the society that align interests in this way. We explored piecemeal approaches to achieving this, as well as a general approach that can be applied to the society as a whole. The general approach involves the establishment of a vertical 'invisible hand' process to complement the horizontal 'invisible hand' of economic markets. A key goal of vertical markets is to develop systems for establishing governance and other institutions that are as innovative, distributed, dynamic, collectively intelligent and self-organizing as are ideal economic markets (we noted that ideal economic markets will not often be achieved in practice without effective vertical markets: if economic markets are to serve the interests of society, they must be well-managed).

Without the forms of organization outlined here, modern human societies will continue to fail to self-organize 'the good'. Instead they will continue to harness most of the wealth and resources of the society to satisfy the needs and goals of only a tiny proportion of the population (e.g. see [63]).

As the history of the twentieth century demonstrates, past attempts to change this outcome have been futile. Even if revolution manages to overthrow existing power structures and elites, in the absence of the kinds of fundamental changes to the organization of society that are envisaged here, self-organization tends to produce similar power structures and elites again. George Orwell's Animal Farm illustrates these processes of self-organization very effectively [64]. The repeated failure of movements that have been directed at building a better world is due largely to their inability to envisage how society can be successfully reorganized so that it henceforth inexorably self-organizes 'the good'.

# 5. Acknowledgements

I gratefully acknowledge the benefit of useful discussions and comments from Mark Roddam, Paul Strutynski and David Richards.

# 6. References

1. Heylighen, F. The science of self-organization and adaptivity. In *The Encyclopedia of Life Support Systems*, EOLSS Publishers: Oxford, UK, 2003.
2. Gershenson, C.; Heylighen, F. When can we call a system self-organizing? In Banzhaf, W., Christaller, T., Dittrich, P., Kim, J. T., Ziegler, J. Eds; *Advances in Artificial Life, 7th European Conference, ECAL 2003 LNAI 2801*, SpringerVerlag: Berlin, DE, 2003, pp. 606-614.
3. Hayek, F. A., *The Road to Serfdom*, 2001 ed.; Routledge Classics: London, UK, 1944.
4. Smith, A. *The Wealth of nations*, 1994 ed.; Modern Library: New York, NY, USA, 1776.
5. Maynard Smith, J.; Szathmary, E. *The Major Transitions in Evolution*, W. H. Freeman: New York, NY, USA, 1995.
6. Stewart, J. E. *Evolution's Arrow*, The Chapman Press: Canberra, AU, 2000.
7. Stewart, J. E. Metaevolution. *J. Soc. Evol. Syst*. **1995**, 18, 113-114.
8. Stewart, J. E. Evolutionary transitions and artificial life. *Artificial Life* **1997**, 3, 101-120.
9. Stewart, J. E. Evolutionary Progress. *J. Soc. Evol. Syst.* **1997**, 20, 335-62.
10. Stewart, J. E. The meaning of life in a developing universe. *Found. Sci.* **2010**, 15, 395-409.
11. Stewart, J. E. The direction of evolution: the rise of cooperative organization. *Biosystems* **2014**, 123, 27-36.
12. Okasha, O. *Evolution and the Levels of Selection*, Clarendon Press: Oxford, UK, 2006.
13. Corning, P. A. *The Synergism Hypothesis: A Theory of Progressive Evolution*, McGraw-Hill: New York, NY, USA, 1983.
14. Dugatkin, L. *Cheating Monkeys and Citizen Bees: The Nature of Cooperation in Animals and Human*, Free Press: New York, NY, USA, 1999.
15. Miller, J. G. *Living Systems*, McGraw-Hill: New York, NY, USA, 1978.
16. Knoll, K.; Bambach, R. Directionality in the history of life: diffusion from the left wall or repeated scaling of the right? *J. Paleo.* **2000**, 26(supplement), 1–14.
17. Archetti, M.; Scheuring, I. Review: Game theory of public goods in one-shot social dilemmas without assortment. *J. Theor. Biol*. **2012**, 299, 9–20.
18. Nowak, M. A. *Evolutionary Dynamics*, Harvard University Press: Cambridge, MA, USA, 2006.
19. Perc, M; Grigolini, P. Collective behaviour and evolutionary games – an introduction. *Chaos Soliton. Fract.* **2013**, 56, 1-5.
20. Perc, M.; Szolnoki, A. Coevolutionary games – a mini review. *BioSystems* **2010**, 99, 109–125.
21. Sigmund, K. *The Calculus of Selfishness*. Princeton University Press: Princeton, NJ, USA, 2010.
22. Boyd, R.; Richerson, P. Solving the Puzzle of Human Cooperation. In *Evolution and Culture*, Levinson, E., Stephen, C., Jaisson, P., Eds.; MIT Press: Cambridge, MA, USA, 2005; pp. 105-132.
23. Buss, L. W. *The Evolution of Individuality*, Princeton University Press: Cambridge, MA, USA, 1987.
24. Olson, M. *The Logic of Collective Action*, Harvard University Press: Cambridge, MA, USA, 1965.
25. Williams, G. C. *Adaptation and Natural Selection*, Princeton University Press: Cambridge, MA, USA, 1966.
26. Trivers, R. L. The evolution of reciprocal altruism. *Q. Rev. Biol*. **1971,** 46, 35–57.
27. Roberts, G. Evolution of direct and indirect reciprocity. *Proc. R. Soc. B*. **2008**, 275, 173–179.
28. Bronstein, J. L. Our current understanding of mutualism. *Quarterly Review of Biology* **1994**, 69, 31–51.
29. Kauffman, S. A. *The Origins of Order: Self-organization and selection in evolution*, Oxford University Press: Oxford, UK, 1993.
30. Hamilton, W. D. The genetical evolution of social behaviour, I & II. *J. Theor. Biol*. **1964**, 7, 1–52.
31. Griffin, A. S.; West, S. A.; Buckling, A. Cooperation and competition in pathogenic bacteria. *Nature* **2004**, 430, 1024–1027.

32. Nowak, M. A.; Tarnita, C. E.; Antal, T. Evolutionary dynamics in structured populations. *Phil. Trans. R. Soc. B* **2010**, 365, 9–30.
33. Gintis, H. Strong Reciprocity and Human Sociality. *J. Theor. Biol.* **2000**, 206, 169–179.
34. Fehr, E; Gächter, S. Altruistic punishment in humans. *Nature* **2002**, 415, 137–140.
35. Bagley, R. J.; Farmer, J. D. Spontaneous Emergence of a Metabolism. In *Artificial Life I*, Langton, C., Taylor, J., Rasmussen, S., Eds.; Addison and Wesley: New York, NY, USA, 1991, pp. 93-141.
36. Eigen, M.; Schuster, P. The Hypercycle. A Principle of Natural Self-Organization. Part A: Emergence of the Hypercycle. *Naturwissenschaften* **1977**, 64, 541–565.
37. Ulanowicz, R. E. *A Third Window: Natural Life Beyond Newton and Darwin*. Templeton Foundation Press: West Conshohocken, PA, USA, 2009.
38. Maynard Smith, J. Hypercycles and the origin of life. *Nature* **1979**, 280, 445-446.
39. Pross, A. *What is Life?* Oxford University Press: Oxford, UK, 2012.
40. Hall, W. P. Physical basis for the emergence of autopoiesis, cognition and knowledge. *Kororoit Institute Working Papers No. 2* **2011**, 1–63, http://kororoit.org/PDFs/WorkingPapers/Hall-Working0002.pdf (accessed 5 July 2015).
41. Heylighen, F.; Beigi, S.; Veloz, T. Chemical organization theory as a universal modeling framework for self-organization, autopoiesis and resilience. *ECCO Working Paper No. 1* **2015**, http://pespmc1.vub.ac.be/papers/COT-ApplicationSurvey.pdf (accessed 8 May 2015).
42. Stewart, J. E. The Evolution of Cooperative Organization and The Origins of Life. Presented at OEE2 Workshop at the 15th International Conference on the Synthesis and Simulation of Living Systems, Cancun, Mexico, 2016. http://ssrn.com/abstract=2811492 (accessed 1 August 2016).
43. Ostrom, E. *Governing the commons: the evolution of institutions for collective action*, Cambridge University Press: Cambridge, MA, USA, 1990.
44. Beer, S. *Brain of the firm*, Allen Lane: London, UK, 1972.
45. North, D. C. *Institutions, Institutional Change, and Economic Performance*, Cambridge University Press: Cambridge, MA, USA, 1990.
46. Salthe, S. N. *Evolving Hierarchical systems*, Columbia University Press: New York, NY, USA, 1985.
47. Wilson, D. S. *Darwin's Cathedral: Evolution, Religion, and the Nature of Society*, University of Chicago Press: Chicago, IL, USA, 2002.
48. Beinhocker, E. D. *The Origin of Wealth: Evolution, Complexity and the Radical Remaking of Economics*, Random House Business Books: London, UK, 2007.
49. Herman, E. S.; Chomsky, N. *Manufacturing Consent*, Pantheon Books: New York, NY, USA, 1988.
50. Chomsky, N. *Understanding Power: The Indispensable Chomsky*, The New Press: New York, NY, USA, 2002.
51. Chomsky, N. *How the World Works*, Soft Skull Press: Berkeley, CA, USA, 2011.
52. McMurtry, J. *The Structure of Marx's World-View*, Princeton University Press: Princeton, NJ, USA, 1978.
53. Bueno de Mesquita, B. *The Logic of Political Survival*, MIT Press: Cambridge, MA, USA, 2002.
54. Plato. *The Republic*, Simon & Brown: Hollywood, FL, USA, 2011.
55. Shayer, M; Wylam, H. The distribution of Piagetian stages of thinking in British middle and secondary school children II: 14-16 year olds and sex differentials. *British Journal of Educational Psychology* **1978,** 48, 62-70.
56. Pintrich, P. R. Implications of psychological research on student learning and college teaching for teacher education. In *Handbook of Research on Teacher Education*, Houston, W. R., Haberman, M., Sikula, J.; Eds.; MacMillan: New York, NY, USA, 1990, pp.926–857.
57. Stewart, J. E. Review of Book: Dialectical Thinking for Integral Leaders: A Primer. *Integral Leadership Review*, August-November Issue **2016,** http://integralleadershipreview.com/14809-14809/, (accessed 1 October 2016).
58. Stewart, J. E. *The Evolutionary Manifesto*, Kindle ed.; The Chapman Press: Canberra, AU, 2012.
59. Last, L. Digital Distributed Democracy (DDD). *ECCO working paper No. 7* **2015,** http://ecco.vub.ac.be/?q=node/21 26/07/2015 (accessed 26 July 2015)
60. Atlee, T. *The Tao of Democracy: using co-intelligence to create a world that works for all*, BookSurge Publishing: Charleston, SC, USA, 2002.

61. Clarke, C. J. Merging and Dissolving Special Districts. *Yale Journal on Regulation*, **2014**, 31, 493-503.
62. Bunzl, J. *The Simultaneous Policy: An Insider's Guide to Saving Humanity and the Planet*, New European Publications: London, UK, 2001.
63. Piketty, T. *Capital in the Twenty-First Century*, Belknap Press: Cambridge, MA, USA, 2014.
64. Orwell, G. *Animal Farm*, Harcourt Brace Jovanovich: New York. NY, USA, 1946.